# Toroidal Plasmonic Nanodimers for Enhanced Near-Infrared Emission in Heterostructured InP Quantum Dots

Arda Gulucu[1,2] and Emre Ozan Polat[1,2,*]

[1]*Department of Physics, Bilkent University, 06800, Ankara, Turkey*

[2]*UNAM - National Nanotechnology Research Center and Institute of Materials Science and Nanotechnology, Bilkent University, 06800, Ankara, Turkey*

*Corresponding author: emre.polat@bilkent.edu.tr*

**Abstract**

Near-infrared (NIR) emitters operating in the 650-900 nm range are highly attractive for imaging and sensing in turbid media; however, cadmium-free InP-based quantum dots (QDs) often suffer from limited brightness due to nonradiative pathways and inefficient photon outcoupling. In particular, heterostructured InP QDs can exhibit band alignments that induce partial spatial separation of charge carriers, leading to reduced electron-hole wavefunction overlap. This modifies intrinsic recombination dynamics and enhances the sensitivity of their emission to the surrounding photonic environment. Here, we investigate silver toroidal plasmonic nanoantenna dimers (Ag TPNDs) through finite-difference-time-domain (FDTD) simulations as a geometry-tunable platform for enhancing NIR emission of heterostructured InP-based QDs. The coupled toroidal geometry supports strongly confined bonding modes that generate intense nanogap hotspots, while its resonance can be systematically tuned through the toroid aspect ratio. By spectrally aligning the antenna response with QD emission bands (675-845 nm), we achieve large Purcell enhancements together with high quantum efficiencies, demonstrating efficient conversion of enhanced decay rates into radiative emission. We further show that nanometer-scale variations in emitter-antenna separation strongly modulate the radiative rates and spectral response. These results establish toroidal plasmonic nanodimers as a topology-driven platform for controlling emission in NIR quantum emitters and for advancing NIR nanophotonic applications.

**Introduction**

NIR light sources and probes play a central role in applications requiring deep optical penetration and reduced scattering in biological media, including in vivo fluorescence imaging, biosensing, and photonic readout in turbid environments [1–3]. Emitters operating in the 650-900 nm range benefit from improved imaging depth and contrast relative to visible wavelengths, motivating the development of efficient NIR fluorophores and engineered photonic environments that enhance their brightness [4,5]. Colloidal QDs are particularly attractive in this context due to their size-tunable emission, broad absorption, and high photostability [6,7]. Among them, cadmium-free InP-based QDs have emerged as promising alternatives for applications where material toxicity and regulatory constraints are critical; however, their NIR emission efficiency is often limited by nonradiative pathways and imperfect photon outcoupling [8–10]. To overcome the drawback, heterostructured InP-based QD architectures are implemented, which can exhibit quasi-type-II or inverted type-I band alignments depending on shell thickness and composition [11–14]. In these systems, partial spatial separation of charge carriers reduces electron-hole wavefunction overlap, leading to modified recombination dynamics characterized by extended lifetimes and reduced intrinsic radiative rates [15,16].

Particularly, ZnSe/InP/ZnS core-shell-shell QDs form a reverse type-I (quasi-type-II) heterostructure, in which electrons are primarily confined within the InP core while holes are partially delocalized into the ZnSe shell [11,17]. This reduced electron-hole overlap modifies intrinsic recombination dynamics and typically leads to longer lifetimes and lower radiative rates. As a result, the photoluminescence (PL) spectrum of these QDs become particularly sensitive to their electromagnetic environment, making them well-suited for external control via photonic and plasmonic structures [18,19].

A powerful strategy to enhance the apparent brightness of nanoscale emitters is to engineer their electromagnetic local density of optical states (LDOS), thereby modifying spontaneous emission rates through the Purcell effect and controlling the balance between radiative and nonradiative decay channels [20–23]. Plasmonic nanoantennas are particularly effective in this regard due to their ability to confine electromagnetic fields well below the diffraction limit, enabling strong decay-rate enhancement across broadband spectral ranges via localized surface plasmon resonances (LSPRs) [24–30]. However, metal-induced ohmic losses often lead to fluorescence quenching, such that large total decay rate enhancement does not necessarily translate into useful photon emission [31,32]. Consequently, maintaining a radiative decay rate over non-radiative loss is essential for practical applications [33,34].

Nanogap dimers are well-known plasmonic architectures that produce extreme near-field confinement and strong emission-rate enhancement by concentrating charge and field within the interparticle gap [35,36]. In such systems, plasmon hybridization leads to bonding and antibonding modes, where the bonding mode typically exhibits strong radiative character under longitudinal excitation [37–39]. Despite their effectiveness, conventional plasmonic dimers are often limited by increased nonradiative losses at small separations [34,40].

In contrast, toroidal geometries introduce a distinct class of plasmonic modes characterized by poloidal current loops and hybridized inner-outer surface excitations. These modes enable topology-driven control over electromagnetic confinement, spectral tunability, and decay-channel redistribution [41–45]. Our recent study has shown that toroidal nanoantennae (TNA) can access regimes where radiative decay dominates over nonradiative loss, allowing efficient conversion of LDOS enhancement into photon emission [43]. When two toroidal elements are coupled in a dimer configuration, strong near-field interaction further enhances field localization within the gap and enables additional tuning of resonance characteristics [41,46].

Here, we numerically investigate Ag TPNDs as a tunable platform for enhancing fluorescence emission from InP-based QDs using FDTD simulations combined with a photoluminescence-rate framework allow independent evaluation of radiative decay modification, non-radiative quenching, Purcell enhancement and distance dependent spectral dynamics. By systematically varying the nanodimer aspect ratio, we demonstrate spectral alignment of the antenna resonance with multiple QD emission bands. A dipole emitter, as a QD, placed within the nanogap experiences strong LDOS enhancement, leading to large Purcell factors while maintaining high radiative decay rates. We further analyze the dependence of radiative enhancement and spectral response on emitter-antenna separation, establishing TPNDs as a robust platform for NIR quantum emitter engineering. These results highlight the potential of toroidal plasmonic dimers as geometry-controlled, low-loss nanoantennas for spectrally programmable light-matter interactions, enabling efficient and selective emission engineering in heterostructured quantum dots.

**Results**

We begin by modeling the PL emission of ZnSe/InP/ZnS QDs. **Figure 1a** presents a schematic illustration of the ZnSe/InP/ZnS core-shell-shell architecture considered in this study. In zero-dimensional heterostructures, the absorption and emission energies are predominantly governed by quantum confinement effect of each consecutive layer, rendering them highly sensitive to particle size, as well as to the material composition and relative thicknesses of the constituent layers [11,17]. **Figure 1b** conceptually illustrates this quantum-confinement framework, where optical excitation generates an electron-hole pair and the resulting PL transition energy shifts with dot size. Following the experimental study by Saobae *et al.* [11], we model four representative QD emitters with normalized PL spectra centered at approximately 675 nm 740 nm 770 nm and 845 nm with corresponding full width at half-maximum (FWHM) values of 174, 145, 129, and 123 nm, respectively. In the electromagnetic simulations, each QD is approximated as an ideal point electric dipole emitter.

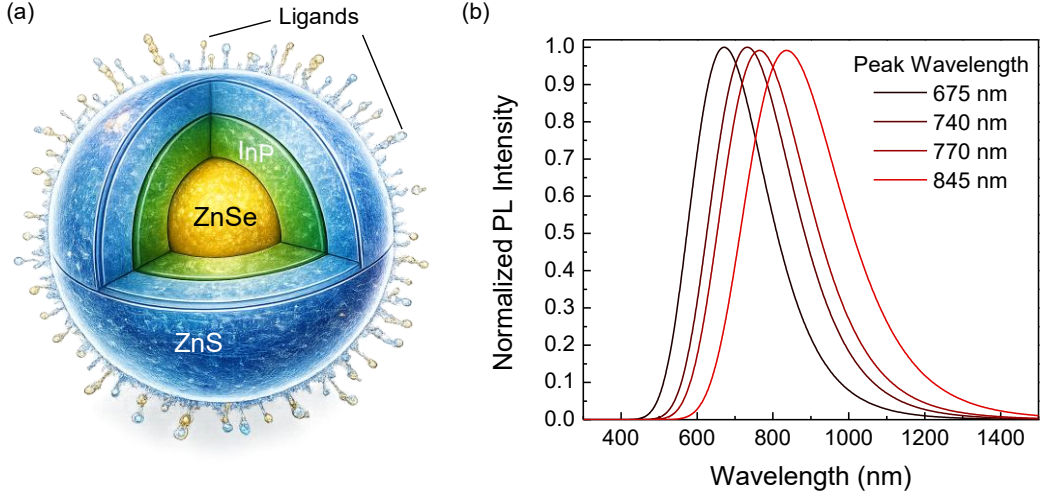

**Figure 1. Normalized photoluminescence spectra of heterostructured QDs with distinct emission maxima (a)** Schematic representation of ZnSe/InP/ZnS core-shell-shell quantum dots functionalized with surface ligands. **(b)** Normalized PL spectra of four QD batches, modeled as point dipole emitters, representing tunable emission peaks centered at approximately 675, 740, 770, and 845 nm.

We next characterize the bare TPNDs under plane-wave excitation, with the incident polarization aligned along the dimer axis (**Figure 2a**). For a representative geometry ($d = 14$ nm, $r + R = 60$ nm), the electric-field intensity enhancement $|E/E_0|^2$ exhibits a strongly confined hotspot within the nanogap (**Figure 2b**), reaching values on the order of $5.4 \times 10^4$ for $r/R = 0.24$. Outside the gap, weaker field localization appears along the toroidal rims, consistent with capacitive coupling between the two antennas. In this regime, opposite charges accumulate across the gap, giving rise to a bonding (in-phase) plasmonic mode with a finite net dipole moment [35].

The scattering and absorption cross sections, extracted via Poynting-flux analysis (**Figure 2c** and **d**), exhibit pronounced and geometry-dependent resonances that can be tuned across the visible-NIR spectral range by varying the toroid aspect ratio $r/R$ [43]. Increasing $r/R$ induces a systematic blue-shift in both scattering and absorption peaks, enabling spectral alignment of the antenna response with targeted QD emission wavelengths. This behavior can be understood in terms of the aspect-ratio-dependent depolarization factor $L(r/R)$, which governs the effective polarizability $\alpha_{\text{eff}}$;

$$\alpha_{\text{eff}}(\omega) = V \frac{\varepsilon - \varepsilon_m}{\varepsilon_m + L(r/R)(\varepsilon - \varepsilon_m)}, \tag{1}$$

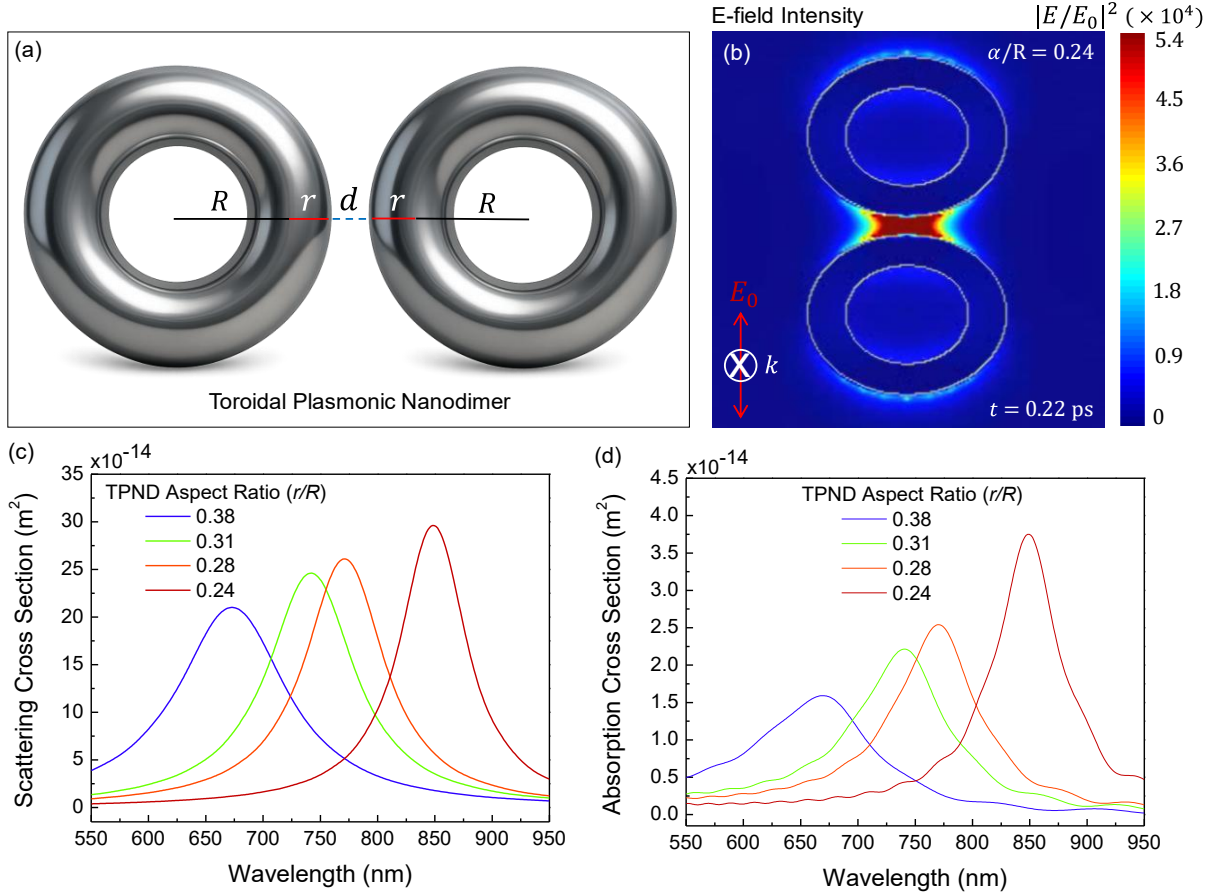

**Figure 2. Toroidal plasmonic nanodimer geometry and aspect-ratio-dependent optical response.** (a) Schematic of a plasmonic dimer composed of two identical Ag TNA separated by a gap of $d = 14$ nm. The geometry is defined by the major radius $R$ and minor radius $r$, with a fixed total size $r + R = 60$ nm; the aspect ratio $r/R$ parametrizes the structure. (b) Simulated electric-field intensity enhancement $\mid E/E_0 \mid^2$ for $r/R = 0.24$ under plane-wave excitation, where the incident polarization $E_0$ is aligned along the dimer axis and the propagation vector $k$ is normal to the plane. A pronounced electromagnetic hotspot forms within the nanogap (color scale $\times 10^4$). (c) Scattering and (d) absorption cross sections as a function of wavelength for fixed gap $d = 14$ nm and varying aspect ratios $r/R = 0.24, 0.28, 0.31$ and $0.38$. The spectra reveal systematic resonance tuning, exhibiting a clear blue-shift with increasing $r/R$ across the visible to NIR spectral range.

where $\omega$ is angular frequency, $V$ is the effective volume of metallic nanostructure, $\varepsilon$ is the complex dielectric function of the metal and $\varepsilon_m$ is the permittivity of the surrounding medium. As $r/R$ increases, the resonance condition shifts according to $\text{Re}\{\varepsilon(\omega_{\text{res}})\} \approx -\frac{1-L}{L}\varepsilon_m$, leading to the observed blue-shift. Conversely, decreasing $r/R$ enhances $\alpha_{\text{eff}}$ and thus the oscillator strength of the bright bonding mode by promoting stronger charge accumulation across the gap, which increases the induced dipole moment. As a result, both scattering ($\propto k^4 \mid \alpha_{\text{eff}} \mid^2$) and absorption ($\propto \text{Im}\{\alpha_{\text{eff}}\}$) cross sections are amplified at resonance. Across all geometries, the plasmonic response remains predominantly radiative, as indicated by the ratios $\frac{\sigma_{\text{sca}}}{\sigma_{\text{abs}}} = 7.90, 10.3, 11.1$ and $13.2$ $r/R = 0.24, 0.28, 0.31$ and $0.38$, respectively, evaluated at their resonance wavelengths. This trend is characteristic of the bright bonding dimer mode, which efficiently couples to far-field radiation.

Next, we place a QD emitter within the dimer gap, oriented along the dimer axis (**Figure 3a**). Based on the comparative analysis of different emitter placements presented in the Supplementraty Information (**Figure S1**), the nanogap configuration was selected for the main study, as it yields the strongest radiative decay-rate enhancement together with the most favorable radiative-to-nonradiative balance. This behavior is consistent with the formation of an intense, hybridized near-field hotspot in the gap region between the two coupled toroidal nanoantennas (TNAs). The emitter position is parameterized by the distance $d_{\text{dip}}$ to the outer surfaces of both TNA; unless otherwise stated, we consider a symmetric configuration with $d_{\text{dip}} = 3$ nm from each TNA. The total decay-rate enhancement (Purcell factor) is evaluated as $\gamma_{\text{tot}}/\gamma_0$, while the radiative component $\gamma_r$ is obtained from far-field power monitors, enabling a consistent extraction of the quantum efficiency $\phi = \gamma_r/(\gamma_r + \gamma_{\text{nr}})$.

To assess performance over emission linewidths, we evaluate both the Purcell factor and radiative decay rate across the full width at half maximum (FWHM) of each QD spectrum (**Figure 3c** and **3d**). In all cases, the enhancement peaks near the emission maximum and remains substantial across the entire bandwidth. The radiative decay rate enhancement closely follows the spectral overlap between the antenna resonance and the QD emission, reaching peak values of 1545, 3033, 2820, and 4602 for QD675, QD740, QD770 and QD845 respectively (**Figure 3d**). These results confirm that toroidal nanodimers can simultaneously deliver large total decay-rate enhancement and strong radiative output across broadband emission profiles. This behavior can be further interpreted using the cavity-like expression for a dipole emitter coupled to a single optical mode, where the Purcell factor scales as:

$$\frac{\gamma_{tot}}{\gamma_0} = \frac{3}{4\pi^2} \frac{Q}{V_{eff}} \left(\frac{\lambda}{n}\right)^3 \qquad (2)$$

As the aspect ratio $r/R$ decreases, the bonding mode redshifts and the electromagnetic field becomes more tightly confined within the gap, effectively reducing $V_{\text{eff}}$ and increasing the Purcell factor. Consequently, for a fixed inter-toroid separation, $\gamma_{\text{tot}}/\gamma_0$ decreases with increasing $r/R$. Importantly, the enhancement remains highly sensitive to the emitter-gap separation, such that even small variations in effective distance can lead to noticeable changes in decay rates, as observed for the intermediate QDs.

Building on the centered emitter configuration discussed in **Figure 3**, we next examine how the dipole-antenna separation influences the coupling dynamics in the TPND system. **Figure 4a** presents the coupling of the QD845 emitter to radiative (photonic) and nonradiative (ohmic) channels, quantified by the decay rates normalized to free-space emission, $\gamma_r/\gamma_0$ and $\gamma_{nr}/\gamma_0$, respectively, evaluated at $d_{\mathrm{dip}} = 3$ and $r/R = 0.24$.

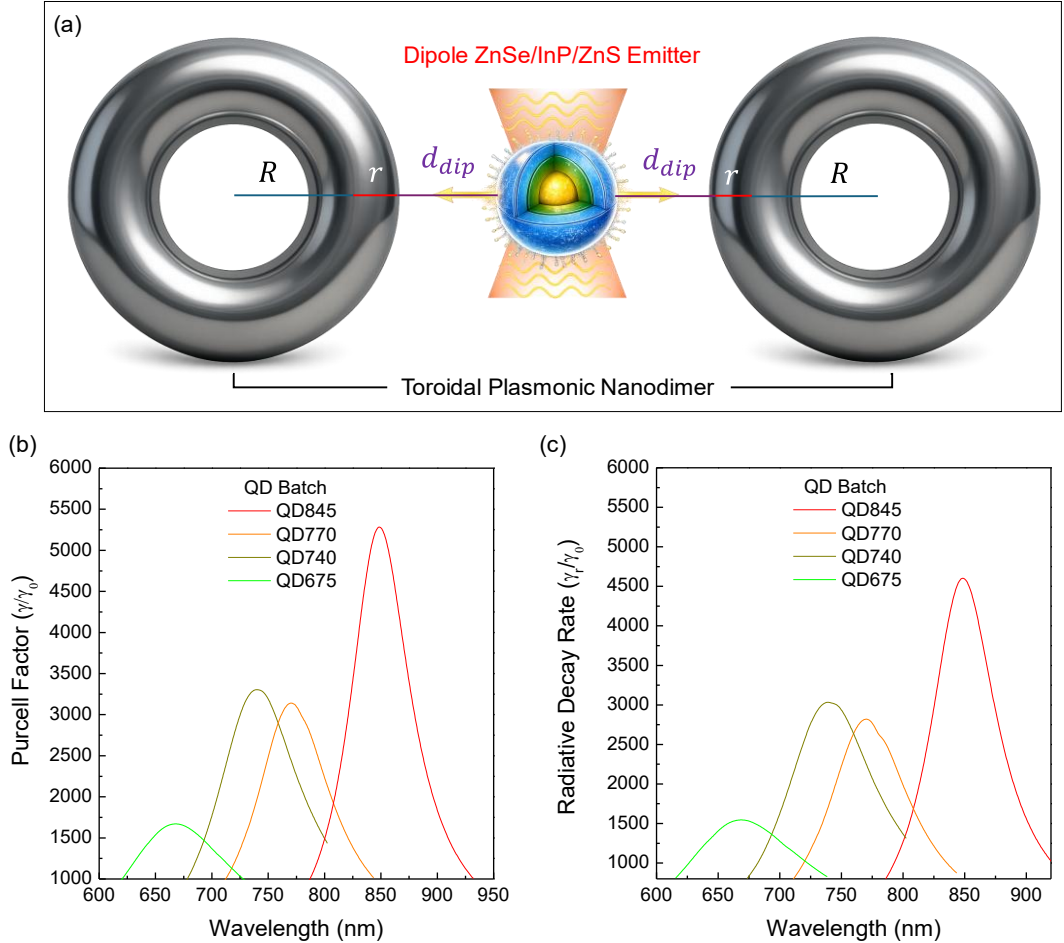

**Figure 3. Toroidal plasmonic nanodimer coupled to a quantum dot emitter.** **(a)** Schematic illustration of a TPND incorporating a ZnSe/InP/ZnS QD emitter, modeled as an electric dipole positioned within the dimer gap and oriented along the dimer axis. The emitter position is defined by the separation distance $d_{\mathrm{dip}}$ from the outer surfaces of the left and right toroids, while the toroidal geometry is modulated by changing the minor radius $r$. **(b)** Spectral dependence of the Purcell factor and **(c)** radiative decay rate ($\gamma_r/\gamma_0$), evaluated across FWHM of the corresponding QD emission bands.

By tuning the TPND aspect ratio such that the dimer resonance spectrally overlaps with each QD emission band, pronounced decay-rate enhancements are achieved at the respective emission peaks (**Figure 4b**). The Purcell factor ranges from 1670 for 675 nm emissive QDs (QD675) to 5281 for 845 nm emissive QDs (QD845). The intermediate range Purcell factors of 3303 and 3140 are recorded for the 740 nm emissive QD (QD740) and 770 nm emissive QDs (QD 770) respectively. The slight non-monotonic trend between QD740 and QD770 arises from geometric considerations: although $d_{\text{dip}}$ is fixed, the finite size of the QDs leads to different effective emitter-antenna separations. In particular, QD770 and QD740 correspond to effective gap sizes of 14 nm and 12.3 nm, respectively. Since the Purcell enhancement increases sharply with decreasing separation, this difference leads to a higher enhancement for QD740 compared to what would be expected based on spectral alignment alone.

Across all emitters, the quantum efficiency remains high, with $\phi = 0.93$ (for QD675), 0.92 (for QD740), 0.90 (for QD770), and 0.87 (for QD845), indicating that the enhanced decay is predominantly radiative. Notably, the largest Purcell enhancement (QD845) coincides with the lowest quantum efficiency, reflecting the expected trade-off between extreme field confinement and increased nonradiative losses. The corresponding optimized aspect ratios are $r/R = 0.43$ (for QD675), 0.32 (for QD740), 0.29 (for QD770), and 0.24 (for QD845), demonstrating that geometric tuning of toroidal aspect ratio enables precise spectral matching across the visible-NIR range.

To investigate how emitter placement governs the TPND-QD interaction [29], we systematically varied the dipole-antenna separation $d_{\text{dip}}$ and extracted, the peak wavelength of the response maximum (**Figure 4c**), and the normalized radiative decay rate $\gamma_r/\gamma_0$ (**Figure 4d**). For all four configurations, $\gamma_r/\gamma_0$ decreases monotonically as $d_{\text{dip}}$ increases from 3 to 7 nm, consistent with the decay of the near field and the resulting reduction in coupling strength. The most strongly enhanced emitter, QD845, exhibits a decrease from 4602 at $d_{\text{dip}} = 3\ nm$ to 1760 at $d_{\text{dip}} = 7\ nm$. The remaining emitters follow the same trend with lower absolute values: QD770 decreases from 2820 to 1069, QD740 from 3033 to 1019, and QD675 from 1545 to 511. Thus, increasing the separation by only a few nanometers reduces the radiative enhancement by approximately a factor of 2-3, highlighting the strong distance sensitivity of the coupling in the few-nanometer regime.

Concurrently, the peak wavelength exhibits a systematic blue shift with increasing $d_{\text{dip}}$. Specifically, the resonance shifts from approximately 848 to 830 nm (QD845), 770 to 753 nm (QD770), 740 to 719 nm (QD740), and 670 to 650 nm (QD675) as the separation increases from 3 to 7 nm. This behavior reflects reduced reactive loading of the plasmonic mode at larger separations: at small distances, strong near-field coupling induces a red-shifted hybrid response, whereas at larger separations the resonance approaches that of the uncoupled antenna.

Overall, this interplay between spectral shifting and decay-rate modulation underscores the central role of near-field coupling in governing light-matter interactions in the dimer system. The controllable interplay between spectral position and emission enhancement provides a versatile route for optimizing nanoscale light sources, particularly for NIR photonic and bioimaging applications.

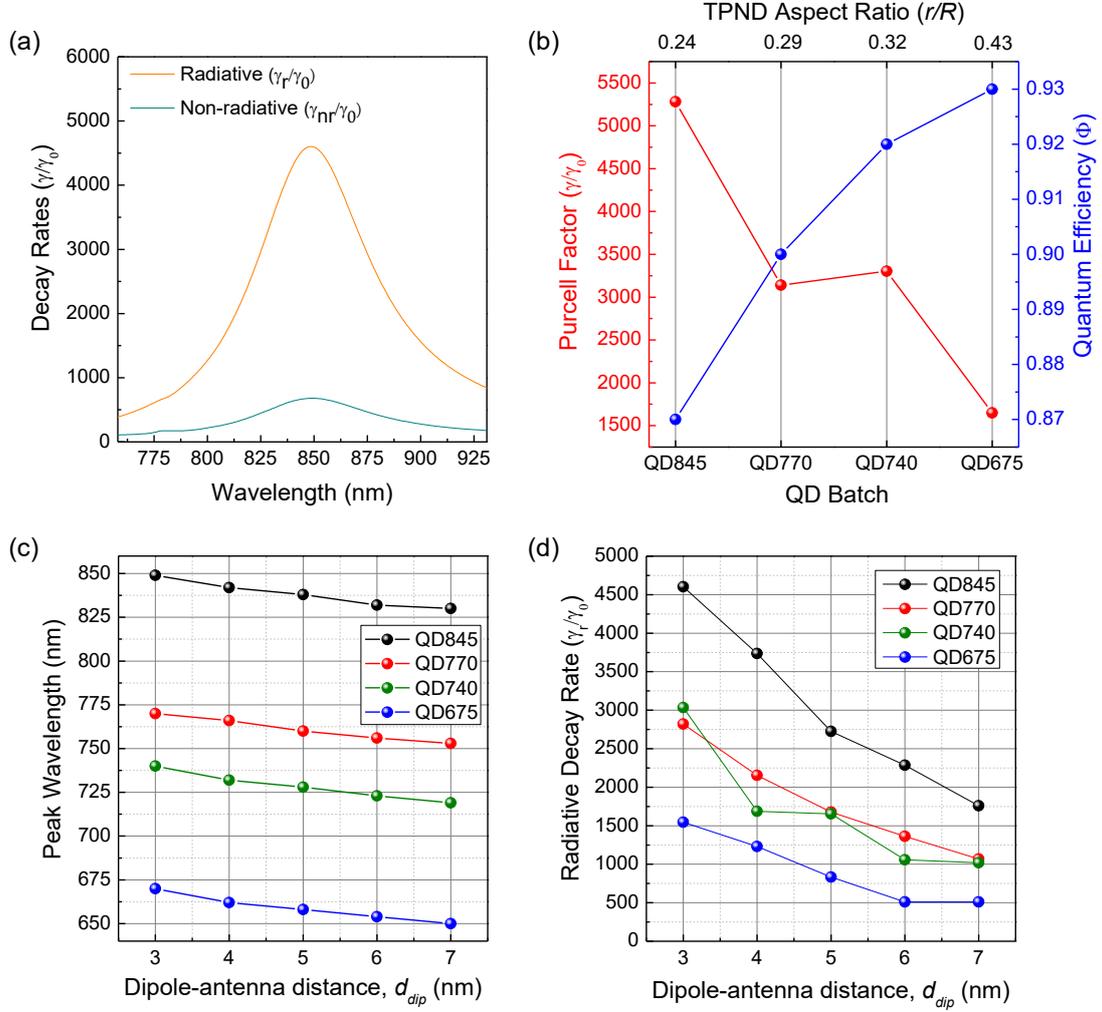

Figure 4. Spectral and distance-dependent radiative and nonradiative decay dynamics of the QD-TPND coupled system. (a) Normalized radiative ($\gamma_r/\gamma_0$) and nonradiative ($\gamma_{nr}/\gamma_0$) decay rates for QD845, evaluated at $d_{\text{dip}} = 3$ nm and $r/R = 0.24$. (b) Purcell factor ($\gamma/\gamma_0$) and quantum efficiency $\phi = \gamma_r/(\gamma_r + \gamma_{nr})$, evaluated at the emission peaks of the four QD batches for a centered emitter configuration. The TPNDs are geometrically tuned such that their resonances spectrally overlap with each QD emission band, maximizing radiative outcoupling. (c) Peak emission wavelength and (d) normalized radiative decay rate ($\gamma_r/\gamma_0$) as a function of dipole-antenna separation $d_{\text{dip}}$. The observed blue shift of the peak wavelength with increasing separation reflects reduced near-field coupling and diminished reactive loading of the plasmonic mode. Concurrently, the monotonic decrease in $\gamma_r/\gamma_0$ highlights the rapid decay of the antenna near field, leading to weakened emitter-antenna coupling at larger distances.

**Conclusion and Discussion**

We demonstrate that Ag TPNDs constitute a compact and geometrically tunable platform for enhancing the emission of InP-based QDs across the NIR spectral range. This enhancement arises from the simultaneous increase in the total decay rate (Purcell factor) and the preservation of high radiative efficiency. Under plane-wave excitation, the dimer supports a strongly confined bonding-mode nanogap hotspot, while its spectral response can be systematically tuned from the visible to the NIR by varying the aspect ratio ($r/R$). Specifically, increasing $r/R$ induces a blue shift in the scattering and absorption resonances, enabling precise spectral alignment with targeted QD emission bands.

The bonding mode remains predominantly radiative across the investigated parameter space, consistent with the large scattering-to-absorption cross-section ratios observed at resonance. When a dipole emitter is positioned within the nanogap and the dimer resonance is tuned accordingly, substantial on-resonance enhancements are achieved. The Purcell factor ranges from 1670 (QD675) to 5281 (QD845), accompanied by consistently high quantum efficiencies ($\phi \approx 0.87$-$0.93$), demonstrating efficient conversion of enhanced LDOS into far-field photon emission. Importantly, this enhancement persists across realistic emission linewidths, with radiative decay-rate enhancements reaching 1545 (QD675), 3033 (QD740), 2820 (QD770), and 4602 (QD845), confirming that brightness enhancement is governed by spectral overlap between the dimer resonance and the quantum dot emission.

These observations follow a cavity-like scaling behavior: decreasing $r/R$ leads to a red shift of the bonding resonance and stronger field confinement (reduced effective mode volume), resulting in increased decay-rate enhancement. Finally, the emission enhancement exhibits strong sensitivity to emitter placement. Increasing the dipole-antenna separation from 3 to 7 nm reduces the radiative decay-rate enhancement by approximately a factor of 2-3 and induces a systematic blue shift in the response, while the quantum efficiency remains comparatively stable.

Beyond these fundamental insights, the operation of the platform in the NIR spectral range directly addresses the challenges of light propagation in turbid media, where scattering and absorption severely limit imaging depth and signal integrity. By operating within the biological optical window, the demonstrated high-brightness, high-efficiency emission is particularly well-suited for applications requiring deep light penetration, such as bioimaging and sensing in scattering environments. The ability of TPNDs to boost emission while maintaining a dominant radiative channel is therefore critical for overcoming photon loss and diffusion in such media.

Overall, these results establish TPNDs as a versatile platform for engineering light-matter interactions at the nanoscale [47], with strong potential for high-efficiency nanoscale emitters [48], deep-tissue optical imaging [49], and integrated photonic and quantum technologies [50].

**Methodology**

Maxwell's curl equations are discretized in both space and time as

$$\nabla \times \mathbf{E} = -\mu \frac{\partial \mathbf{H}}{\partial t}, \nabla \times \mathbf{H} = \varepsilon \frac{\partial \mathbf{E}}{\partial t} + \mathbf{J}, \tag{3}$$

where $\mathbf{E}$ and $\mathbf{H}$ denote the electric and magnetic fields, respectively, $\mu$ is the permeability, $\varepsilon$ is the permittivity, and $\mathbf{J}$ is the current density. The equations are solved in three dimensions using the FDTD method, enabling full-wave simulations of the electromagnetic response of the system in free space. All simulations were performed using a three-dimensional FDTD solver (Ansys Lumerical FDTD) with spatial and temporal discretization chosen to satisfy the Courant stability condition.

Open-boundary conditions are applied as perfectly matched layers (PMLs) to suppress artificial reflections. The toroidal nanoantenna geometry is constructed using Lumerical's torus script, which approximates a torus by discretizing the azimuthal angle into $N$ segments and placing $N$ overlapping extruded circular elements (short cylinders) along a circular trajectory. A value of $N = 400$ was selected after verifying convergence of the optical response. The dielectric function of Ag is taken from the experimentally measured data of Johnson and Christy [51].

The scattering and absorption cross sections are obtained by normalizing the corresponding power quantities to the incident intensity. A total-field/scattered-field (TFSF) plane-wave source is used to define the incident field, with incident intensity $I_{\text{inc}}(\omega)$. The scattered power is calculated from the net outward flux of the time-averaged Poynting vector through a closed surface $S_{\text{out}}$ enclosing the scatterer in the scattered-field region,

$$P_{\text{scat}}(\omega) = \oint_{S_{\text{out}}} \langle \mathbf{S}_{\text{scat}}(\omega) \rangle \cdot d\mathbf{A}, \tag{4}$$

where $\langle \mathbf{S} \rangle = \frac{1}{2}\text{Re}\{\mathbf{E} \times \mathbf{H}^*\}$ is the time-averaged Poynting vector. The absorbed power is computed as the negative net total-field flux through a closed surface $S_{\text{in}}$ enclosing the structure in the total-field region,

$$P_{\text{abs}}(\omega) = -\oint_{S_{\text{in}}} \langle \mathbf{S}_{\text{tot}}(\omega) \rangle \cdot d\mathbf{A}. \tag{5}$$

The absorption and scattering cross sections are then given by

$$\sigma_{\text{abs}}(\omega) = \frac{P_{\text{abs}}(\omega)}{I_{\text{inc}}(\omega)}, \qquad \sigma_{\text{sca}}(\omega) = \frac{P_{\text{scat}}(\omega)}{I_{\text{inc}}(\omega)}. \tag{6}$$

The quantum emitter is modeled as an ideal point electric dipole. The spontaneous emission rate in free space is given by

$$\gamma_0 = \frac{\omega^3 |\mathbf{p}|^2}{3\pi\varepsilon_0 \hbar c^3}, \tag{7}$$

where $\omega$ is the angular frequency, $|\mathbf{p}|$ is the dipole moment, $\varepsilon_0$ is the vacuum permittivity, $\hbar$ is the reduced Planck constant, and $c$ is the speed of light.

The radiative decay rate is obtained from the far-field radiated power,

$$\gamma_r = \frac{P_{\text{rad}}}{P_0} \gamma_0, \tag{8}$$

where $P_{\text{rad}}$ is the power collected by far-field monitors enclosing the emitter-nanodimer system, and $P_0$ is the radiated power of the dipole in free space. The total decay rate is similarly defined as

$$\gamma_{\text{tot}} = \frac{P_{\text{dip}}}{P_0} \gamma_0, \tag{9}$$

where $P_{\text{dip}}$ is the total emitted power in the presence of the nanodimer, obtained from near-field monitors. The nonradiative decay rate is then calculated as

$$\gamma_{nr} = \gamma_{\text{tot}} - \gamma_r. \tag{10}$$

To ensure numerical accuracy, a refined spatial mesh with element sizes ranging from $\lambda/2000$ to $\lambda/1000$ is employed in the near-field region for different $d_{\text{dip}}$ values. Near- and far-field power monitors are strategically positioned to fully enclose the emitting dipole and the dipole–nanodimer system, ensuring accurate capture of all energy channels, including those modified by near-field coupling.

**Acknowledgments**

E.O.P. acknowledges financial support from the Scientific and Technological Research Council of Türkiye (TÜBİTAK) through Grant No. 222N308 within the CHIST-ERA program. Additional support was provided by the Turkish Academy of Sciences Outstanding Young Scientists Awards(GEBİP) 2025.

**Author Contributions**

A.G. performed the numerical simulations, carried out the calculations, and extracted the raw data underlying the reported results. E.O.P. conceived the research idea, supervised the project, and led the preparation of the manuscript, including the development of the final figures and the main text. All authors discussed the results and contributed to the final version of the manuscript.

**Competing interests**

The authors declare no competing interests.

**Data availability**

All the data needed to evaluate the shown results is present within the paper. Additional data related to the paper that support the findings of this study are available from the corresponding author upon reasonable requests.

# Supplementary Information

## for

## Toroidal Plasmonic Nanodimers for Enhanced Near-Infrared Emission in ZnS/InP/ZnSe Quantum Dots

Arda Gulucu[1,2] and Emre Ozan Polat[1,2,*]

[1]*Department of Physics, Bilkent University, 06800, Ankara, Turkey*

[2]*UNAM - National Nanotechnology Research Center and Institute of Materials Science and Nanotechnology, Bilkent University, 06800, Ankara, Turkey*

*\*Corresponding author: emre.polat@bilkent.edu.tr*

To validate the choice of emitter placement used throughout the main analysis, we systematically investigated the dependence of radiative and nonradiative decay dynamics on different spatial configurations of the quantum emitter relative to the toroidal nanoantenna system. As shown in **Figure S1**, three representative configurations were considered: (i) a single toroidal nanoantenna with the emitter positioned at its center, (ii) a side-coupled configuration with the emitter located near a single toroid in a dimer geometry, and (iii) the nanogap configuration, where the emitter is placed between two coupled toroidal nanoantennas. This comparative analysis is essential to disentangle the role of near-field coupling and plasmon hybridization in governing emission enhancement. The results demonstrate that the nanogap configuration provides significantly stronger radiative decay-rate enhancement while maintaining a favorable balance between radiative and nonradiative channels, thereby justifying its selection as the primary configuration in the main text.

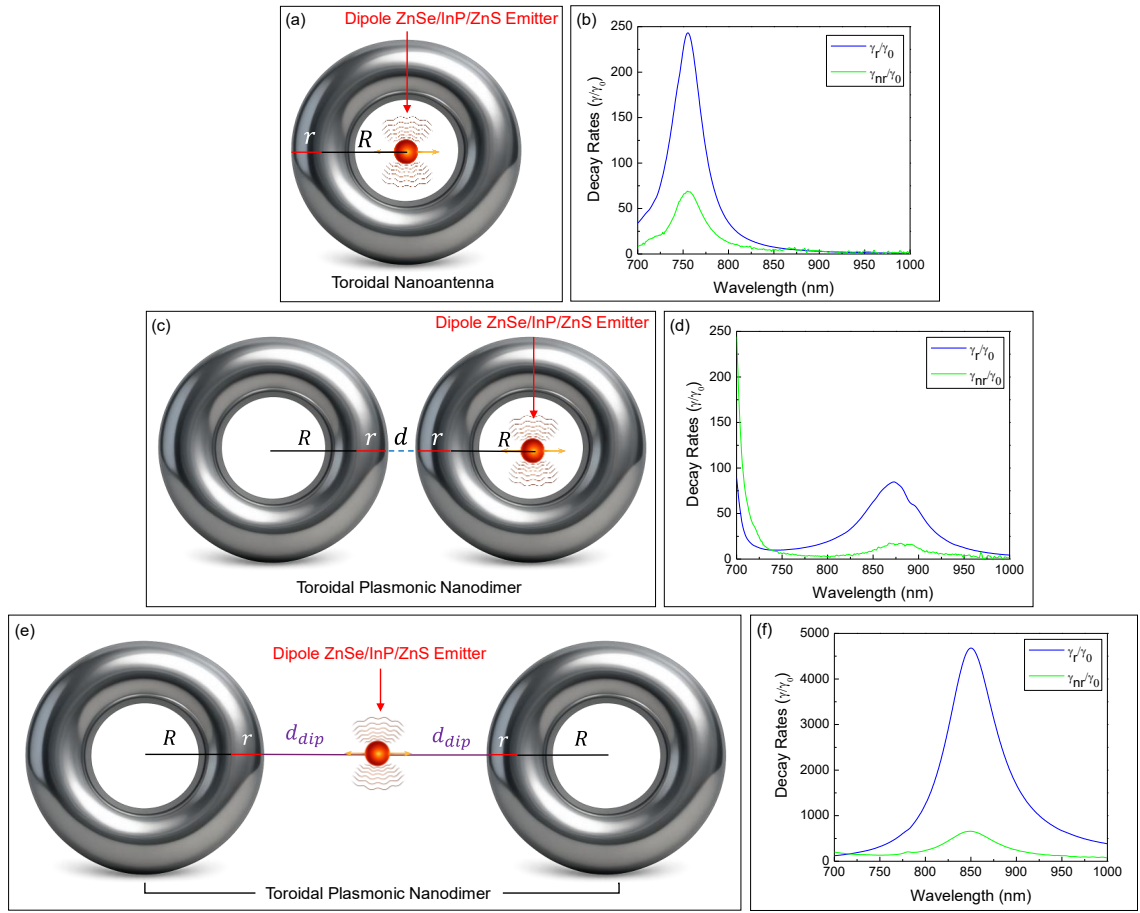

**Figure S1. Comparison of emitter placement configurations in toroidal nanoantenna systems.** **(a)** Schematic illustration of a quantum emitter positioned at the center of a single toroidal nanoantenna. **(b)** Corresponding normalized radiative ($\gamma_r/\gamma_0$) and nonradiative ($\gamma_{nr}/\gamma_0$) decay rates as a function of wavelength, showing moderate radiative enhancement accompanied by a noticeable nonradiative contribution due to limited field confinement. **(c)** Schematic illustration of a side-coupled configuration, where the emitter is positioned near one toroid in a dimer system. **(d)** Corresponding normalized radiative ($\gamma_r/\gamma_0$) and nonradiative ($\gamma_{nr}/\gamma_0$) decay rates, indicating improved coupling compared to the single-toroid case, but still constrained by asymmetric field localization and increased dissipative losses. **(e)** Schematic illustration of the nanogap configuration, where the emitter is placed between two coupled toroidal nanoantennas. **(f)** Corresponding normalized radiative ($\gamma_r/\gamma_0$) and nonradiative ($\gamma_{nr}/\gamma_0$) decay rates, demonstrating a pronounced radiative enhancement with a comparatively lower relative nonradiative contribution due to the strongly confined hybridized hotspot in the gap region. The nanogap configuration exhibits the strongest radiative enhancement and the most favorable radiative-to-nonradiative ratio, highlighting the role of hybridized near-field coupling in the dimer gap.